\documentclass{svmult}

\usepackage{microtype}

\usepackage{multirow}

\usepackage{amssymb}
\usepackage{pifont}
\newcommand{\cmark}{\ding{51}}%
\newcommand{\xmark}{\ding{55}}%

\usepackage{wrapfig}

\usepackage[utf8]{inputenc}

\usepackage[T1]{fontenc}
\usepackage{amsmath}
\usepackage{bm}
\usepackage{xspace}

\usepackage[english]{babel}

\usepackage[bookmarks=false]{hyperref} 

\usepackage{todonotes}

\usepackage{siunitx}

\usepackage{subfig}
\usepackage[export]{adjustbox}

\usepackage{dblfloatfix}

\usepackage{blindtext}

\usepackage{listings}

\usepackage{xcolor}

\newcommand{\orcid}[1]{\href{https://orcid.org/#1}{\includegraphics[height=2ex]{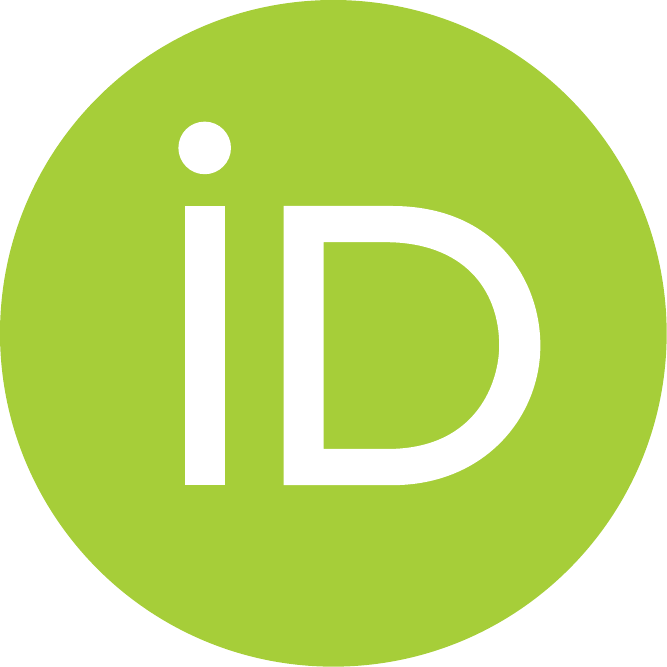}}}

\title*{Advanced Python Performance Monitoring with Score-P}

\author{Andreas Gocht\orcid{0000-0003-2760-3672} \and Robert Sch{\"o}ne \and Jan Frenzel}

\institute{Andreas Gocht \at Center for Information Services and High Performance Computing (ZIH), Technische Universit\"{a}t Dresden, 01062 Dresden, Germany \email{andreas.gocht@tu-dresden.de}
\and Robert Sch{\"o}ne \at Center for Information Services and High Performance Computing (ZIH), Technische Universit\"{a}t Dresden, 01062 Dresden, Germany \email{robert.schoene@tu-dresden.de}
\and Jan Frenzel \at Center for Information Services and High Performance Computing (ZIH), Technische Universit\"{a}t Dresden, 01062 Dresden, Germany \email{jan.frenzel@tu-dresden.de}
}

\begin{document}


\ifx\DraftModeOn\undefined

\newcommand{\todoti}[1]{}
\newcommand{\todomb}[1]{}
\newcommand{\todors}[1]{}
\newcommand{\todoag}[1]{}
\newcommand{\tododh}[1]{}

\newcommand{\figref}[1]{Figure~\ref{fig:#1}}
\newcommand{\tabref}[1]{Table~\ref{tab:#1}}
\newcommand{\secref}[1]{Section~\ref{sec:#1}}
\newcommand{\lstref}[1]{Listing~\ref{lst:#1}}

\newcommand{\papertarget}[2]{}

\renewcommand{\todo}[1]{}

\else

\newcommand{\todoti}[1]{\todo[color=yellow!60,inline,size=\small]{Thomas: #1}}
\newcommand{\todomb}[1]{\todo[color=cyan!60,inline,size=\small]{Mario: #1}}
\newcommand{\todors}[1]{\todo[color=green!60,inline,size=\small]{Robert: #1}}
\newcommand{\todoag}[1]{\todo[color=orange!60,inline,size=\small]{Andreas: #1}}
\newcommand{\tododh}[1]{\todo[color=red!60,inline,size=\small]{Daniel: #1}}

\newcommand{\figref}[1]{\textcolor{red}{Figure~\ref{fig:#1}}}
\newcommand{\tabref}[1]{\textcolor{red}{Table~\ref{tab:#1}}}
\newcommand{\secref}[1]{\textcolor{red}{Section~\ref{sec:#1}}}
\newcommand{\lstref}[1]{\textcolor{red}{Listing~\ref{lst:#1}}}

\newcommand{\papertarget}[2]{\todo[inline]{Target: #1: \\Deadline: #2}}

\fi

\newcommand{\settrace}[0]{\texttt{sys.settrace()}\xspace}
\newcommand{\setprofile}[0]{\texttt{sys.setprofile()}\xspace}
\newcommand{\none}[0]{\texttt{None}\xspace}
\newcommand{\execve}[0]{\texttt{os.execve()}\xspace}
\newcommand{\ldpreload}[0]{\texttt{LD\_PRELOAD}\xspace}
\newcommand{\scorep}[0]{\mbox{Score-P}\xspace}
\newcommand{\bindings}{\scorep \mbox{C-bindings}\xspace}

\definecolor{codegreen}{rgb}{0,0.6,0}
\definecolor{codegray}{rgb}{0.5,0.5,0.5}
\definecolor{codeblue}{rgb}{0,0,0.3}
\definecolor{codepurple}{rgb}{0.58,0,0.82}

\lstdefinestyle{mystyle}{ 
	commentstyle=\color{codegreen},
	keywordstyle=\color{codeblue},
	numberstyle=\tiny\color{codegray},
	stringstyle=\color{codepurple},
	basicstyle=\footnotesize\ttfamily,
	breakatwhitespace=false,         
	breaklines=true,                 
	captionpos=b,                    
	keepspaces=true,                 
	numbers=left,                    
	numbersep=5pt,                  
	showspaces=false,                
	showstringspaces=false,
	showtabs=false,                  
	tabsize=2
}
\lstset{style=mystyle}

\maketitle


\abstract{Within the last years, Python became more prominent in the scientific community and is now used for simulations, machine learning, and data analysis.
All these tasks profit from additional compute power offered by parallelism and offloading.
In the domain of High Performance Computing (HPC), we can look back to decades of experience exploiting different levels of parallelism on the core, node or inter-node level, as well as utilising accelerators.
By using performance analysis tools to investigate all these levels of parallelism, we can tune applications for unprecedented performance.
Unfortunately, standard Python performance analysis tools cannot cope with highly parallel programs.
Since the development of such software is complex and error-prone, we demonstrate an easy-to-use solution based on an existing tool infrastructure for performance analysis.
In this paper, we describe how to apply the established instrumentation framework \scorep to trace Python applications.
We finish with a study of the overhead that users can expect for instrumenting their applications.}

\section{Introduction}
\label{sec:intro}
Python is one of the Top 5 programming languages\footnote{According to the TIOBE Index Oktober 2019: https://www.tiobe.com/tiobe-index/}, and it is not surprising that more and more scientific software is written in Python.
But the standard implementation CPython interprets Python source code, rather than compiling it.
Hence, it is deemed to be less performant than other programming languages like C or C++.
Moreover, as CPython employs a Global Interpreter Lock (GIL)~\cite{2019_glossary}, it is often stated that Python does not support parallelism.
While there are different Python implementations like pypy\footnote{https://pypy.org/} or IronPython\footnote{https://ironpython.net/}, which try to counter these drawbacks, these approaches do not represent the standard implementation.

However, CPython is easily extensible, e.g., by using its C-API or foreign function interfaces.
These interfaces allow programmers to exploit the parallelism of a problem with traditional programming languages like C without losing the flexibility and the power of the standard Python implementation.
Moreover, it is possible to offload computation to accelerators like graphic cards.
Nevertheless, these extensions and the Python source code itself need to be optimised to exploit the full performance of a computing system.
To optimize the application, it has to be monitored.
To monitor the application, performance-related information has to be collected and recorded.

While collecting performance information is possible to some extent with tools that are part of the standard Python installation, none of these tools makes it easy to gain knowledge about the efficiency of thread parallel, process parallel, and accelerator-supported workloads.
However, such tools exist for traditional programming languages used in High Performance Computing (HPC).
Here, \scorep~\cite{Knuepfer2012}, Extrae~\cite{Wagner2017}, TAU~\cite{Shende2006}, and others allow users to record the performance of their applications and analyze them with scalable interfaces.

In this paper, we present the Python bindings for \scorep, which make it easy for users to trace and profile\footnote{As defined in  \cite[Section 2]{2015_ilsche_tracing}} their Python applications, including the usage of (multi-threaded) libraries, MPI parallelism and accelerator usage.
The paper is structured as follows:
We describe our concept and implementation in \secref{concept} and evaluate the overhead in \secref{evaluation}.
We present related work in \secref{related} and finalize this paper with a conclusion and an outlook in \secref{summary}. 

\section{The \scorep Python Bindings}
\label{sec:concept}
\begin{figure}[htb]
	\centering
	\includegraphics[width=.71\columnwidth]{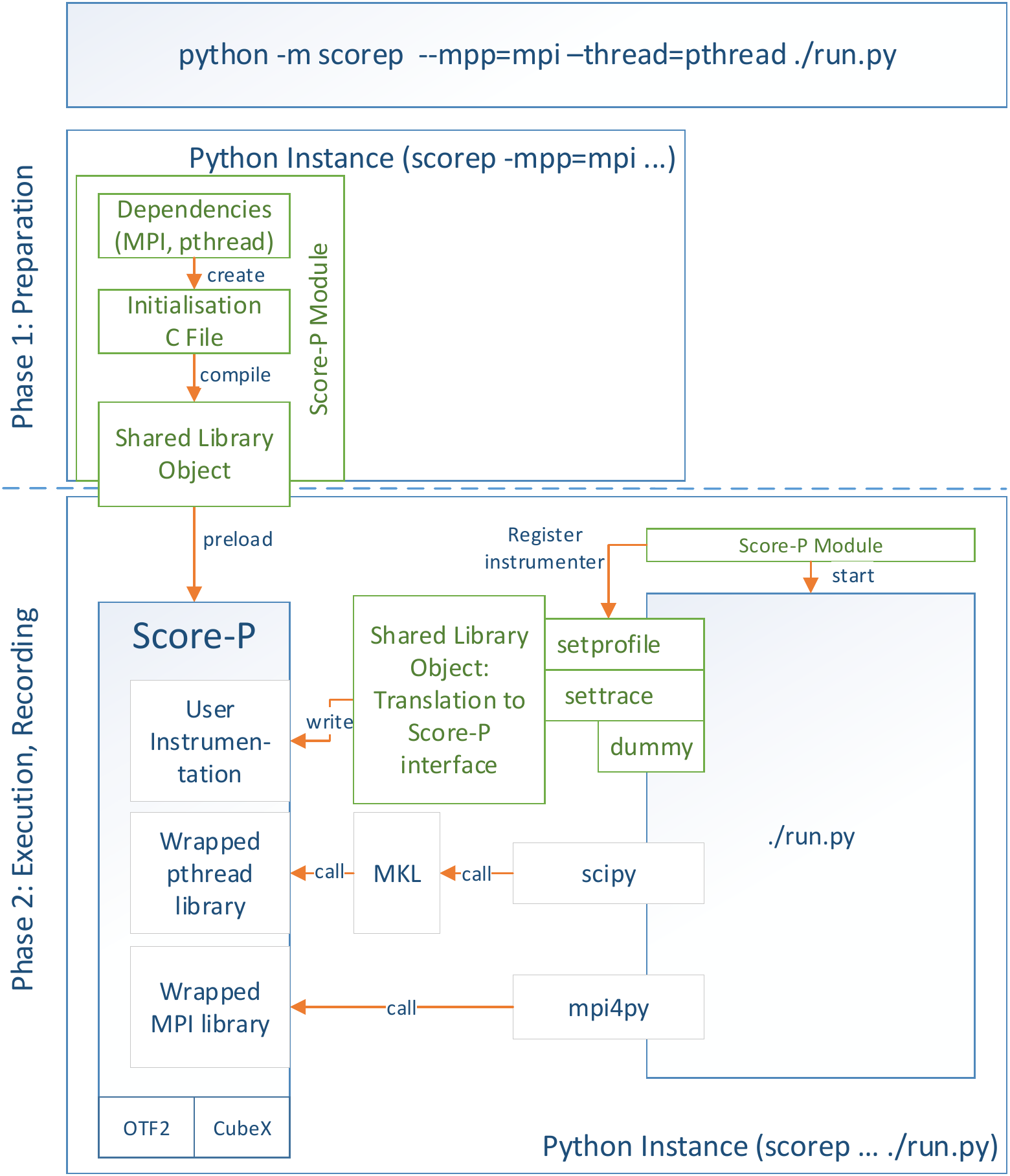}
	\caption{ 
		Overview of the instrumentation process with \scorep.
		In the first phase, the \scorep Python module initializes the \scorep measurement system and attaches \scorep libraries.
		In the second phase, the bindings use the preloaded \scorep libraries and instrument the Python code to record events with \scorep.
		In addition to the Python instrumentation, other parts of the application, such as MPI, pthreads, and CUDA functions, are automatically instrumented by \scorep (not depicted).
	}
	\label{fig:overview}       
\end{figure}

The Python module, which is used to invoke \scorep and allows tracing and profiling of Python code, is called \textit{\scorep Python bindings}.
The module can be split into three basic blocks, which are used in two phases:
The \textit{initialisation}, which is executed in a preparation phase, prepares the measurement and executes the application.
The \textit{instrumenter} is registered with the Python instrumentation hooks and used during execution.
The \textit{\bindings} connect Python with C and Score-P and are also used during execution.
The workflow of the overall process, including preparation phase and the execution phase, is depicted in \figref{overview}.

\subsection{Preparation Phase}
\label{sec:initalisation}
Since version 2.5, Python allows running modules as scripts~\cite{Coghlan2004}.
This approach can be used to record traces of a Python application.
Instead of starting the Python application directly, the script and its parameters are passed as arguments to the \scorep Python module.
The recording can be configured by prefixing additional parameters to the parameter specifying the original Python application.
An example is given in \lstref{scorep_module}.

In the first step, all \scorep related parameters are parsed.
\scorep supports a variety of programming models like OpenMP, MPI, and CUDA.
However, increasing the monitoring detail leads to more information in a trace or profile but also to a higher instrumentation overhead.
Therefore, we allow the user to choose which functionality should be monitored.
Based on the chosen features, the \scorep initialisation code is generated.
This code is then compiled and added together with some dependencies to the \ldpreload environment variable.
As \ldpreload is evaluated by the linker, the whole Python interpreter needs to be restarted, which is done using \execve~\cite{2019_os}.

Once restarted, the module starts the second step: 
The instrumenter is created, and the arguments, which are succeeding the \scorep arguments are utilised.
The first non-\scorep argument is the Python application that shall be executed, followed by its arguments.
The Python application is read, compiled, and executed~\cite{2019_python_built_in}, and its arguments are passed to the application.

\begin{lstlisting}[language=bash, caption=Calling an MPI-parallel application using the \scorep Python bindings, label=lst:scorep_module, frame=single, float=t] 
# mpirun -n 2        -> run two parallel MPI processes
# python             -> each of these runs python
# -m scorep          -> run 'scorep' module before actual script
# --mpp=mpi --thr... -> use MPI & pthread instrumentation
# ./run.py           -> the script or application to run
# -app-arg           -> an argument to ./run.py
mpirun -n 2 \
python -m scorep --mpp=mpi --thread=pthread ./run.py -app-arg
\end{lstlisting}

\subsection{Execution Phase}
\label{sec:instrumenter}
As described before, the execution phase uses two different software parts: the instrumenter and the \bindings that hand over the events from the instrumenter to \scorep.

\subsubsection*{The Instrumenter}
The instrumenter represents a component that is registered with CPython and supposed to be called for specific events during the execution of an application.
Python offers two registration alternatives for such callback functions: \settrace and \setprofile~\cite{2019_python_sys}.
However, different events are raised and forwarded to the instrumenter depending on which of these functions is used.
A summary of these events is shown \tabref{python_events}.
Obviously, both functions can be used to instrument function calls, but both also offer different functionality.
While \setprofile can be used to trace also calls to C-functions, \settrace can be used to record lines of code or operations executed.

Please note that \textit{tracing} has different meanings in the Python documentation and in the HPC community.
In the former, tracing describes the investigation of per line execution of the source code, which can be used to implement debuggers~\cite{2019_python_sys}.
In contrast, the HPC community understands tracing as the recording of events like entering or exiting a region over time~\cite{2015_ilsche_tracing}.
In this paper, we use the term tracing for the HPC notion of tracing.
If we refer to the python notion of tracing we use \settrace.

However, for each callback, \settrace and \setprofile, Python also issues the Python frame causing the event and some additional arguments.
The Python frame holds information like the current line number of the associated module.
The instrumenter passes this information to the \bindings.


\begin{table}[htb]
    \caption{\label{tab:python_events}Supported events for Python profiling/debugging interfaces.}
    \begin{tabular}{llcc}
        \hline
        \multirow{ 2}{*}{Event} & \multirow{ 2}{*}{Description} & \multicolumn{2}{c}{Supported by \texttt{sys\_set\dots}} \\
        &	& \texttt{\dots profile()} & \texttt{\dots trace()} \\
        \hline
        \textit{call} & A function is called & \cmark & \cmark \\
        \textit{return} & A code block (e.g., a function) is about to return & \cmark & \cmark  \\
        \textit{c\_call} & A C function is about to be called & \cmark & \xmark \\
        \textit{c\_return} & A C function has returned & \cmark & \xmark \\
        \textit{c\_exception} & A C function has raised an exception & \cmark & \xmark  \\
        \multirow{ 2}{*}{\textit{line}} & The interpreter is about to a new line  & \multirow{ 2}{*}{\xmark} & \multirow{ 2}{*}{\cmark}\\ 
        & of code or re-execute the condition of a loop \\
        \textit{exception} & An [Python] exception has occurred & \xmark & \cmark \\
        \textit{opcode} & The interpreter is about to execute a new opcode & \xmark  & \cmark \\
            \hline
    \end{tabular}
\end{table}


%
%
%

\subsubsection*{\bindings}
\begin{figure}[b]
		\vspace{1cm}
		\begin{minipage}{.58\textwidth}
			\centering
			\includegraphics[width=\textwidth]{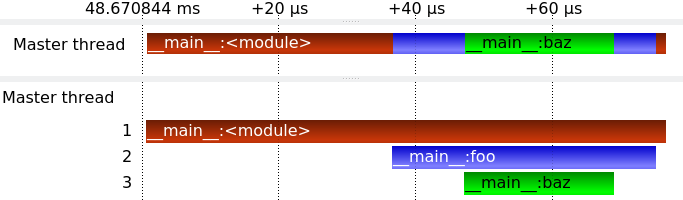}
			\caption{ 
 			   Trace of a simple application using the \scorep Python bindings and Vampir. \texttt{\_\_main\_\_} indicates that the function is part of the currently run script.
 		   }
			\label{fig:vampir1}       
		\end{minipage}
			\hfill
		\begin{minipage}{.38\textwidth}
			\begin{lstlisting}[language=python, caption=Simple Python example, captionpos=b,label=lst:example,frame=single]
def baz():
   print("Hello World")
def foo():
   baz()
if __name__ == \
     "__main__":
   foo()
			\end{lstlisting}
		\end{minipage}
\end{figure}

\begin{figure}[htb]
	\centering 
	\includegraphics[width=\columnwidth]{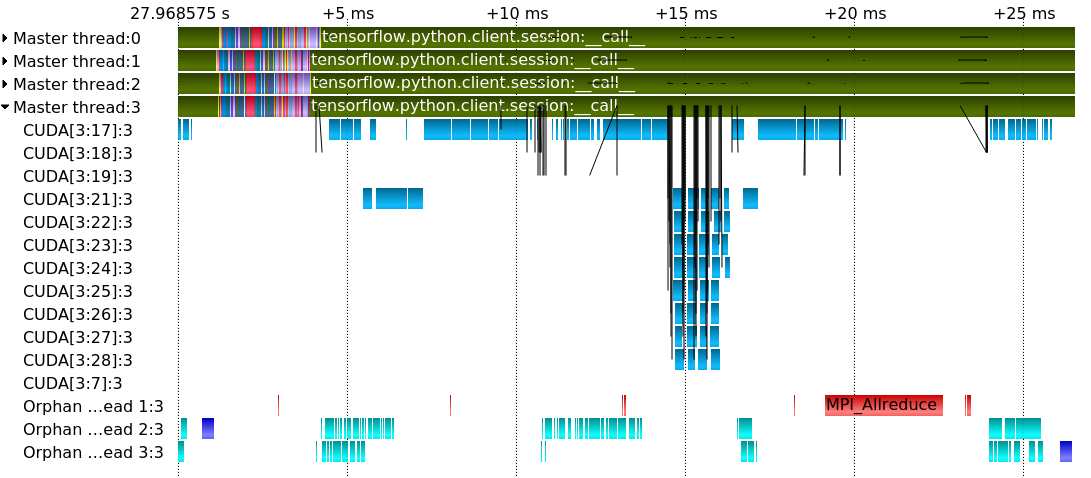}
	\caption{ 
		Trace of a Python application~\cite{2019_horovod} using CUDA and MPI.
		Traced using the \scorep Python bindings.
		Green are TensorFlow functions; red are MPI operations; blue are CUDA operations; black lines are CUDA communication.
	}
	\label{fig:vampir3}       
\end{figure}

The \textit{\bindings} between Python and \scorep use the Python C-interface~\cite{2019_python_ext} and the user instrumentation from \scorep~\cite[Section J.1.2]{2019_scorep}.
The bindings do not only forward events regarding entering or exiting of functions, but also group these functions based on their associated module.
Moreover, they also pass information like line number or the path to the source file to \scorep.
\scorep then uses these instrumentation events to create Cube4-profiles, OTF2-traces or to call substrate plugins for an online interpretation.
Resulting traces can be viewed in Vampir~\cite{vampir}, as shown in \figref{vampir1} for the small example code in \lstref{example}.
A more complex parallel application is visualized in \figref{vampir3}.

\section{Performance Evaluation}
\label{sec:evaluation}
\todoag{[AG] das bleibt wohl drin, das pyHPC paper wird hier die Analyse mit Keras machen.
Vielleicht sollte man hier ein wenig mehr umstellen auf ``es gibt sys.profile und sys.trace, wir messen welches die bessere Idee ist''}

To evaluate the overhead caused by the instrumentation, we designed two test cases.
The first test case, shown in \lstref{test_1}, increments a value in a loop.
We expect that the overhead introduced by the \setprofile instrumenter does not depend on the number of iterations about this loop, since no functions are entered or exited.
In contrast, we expect that the instrumenter using \settrace causes an overhead depending on the iterations, since it is called for each executed line.

The second test case (\lstref{test_2}) uses a function to increment the value.
Here, we expect a strong dependency on the number of iterations for both instrumenters.
\begin{figure}[hb]
\begin{minipage}{.47\textwidth}
\begin{lstlisting}[language=python, label=lst:test_1, caption=Test case 1: loop only,frame=single]
import sys

result = 0

iterations = \
    int(sys.argv[1])

iteration_list = \
    list(range(iterations))

for i in iteration_list:
    result += 1

assert(result == iterations)
\end{lstlisting}
\end{minipage}
\hfill
\begin{minipage}{.48\textwidth}
\begin{lstlisting}[language=python, label=lst:test_2, caption=Test case 2: function calls,frame=single]
import sys

def add(val):
    return val + 1

result = 0
iterations = int(sys.argv[1])
iteration_list = \
    list(range(iterations))

for i in iteration_list:
    result = add(result)
    
assert(result == iterations)
\end{lstlisting}
\end{minipage}
\end{figure}

\begin{figure}[t]
	\centering
	\subfloat[Test case 1 (\label{fig:test_case_1}\lstref{test_1})]{\includegraphics[width=.49\columnwidth,valign=t]{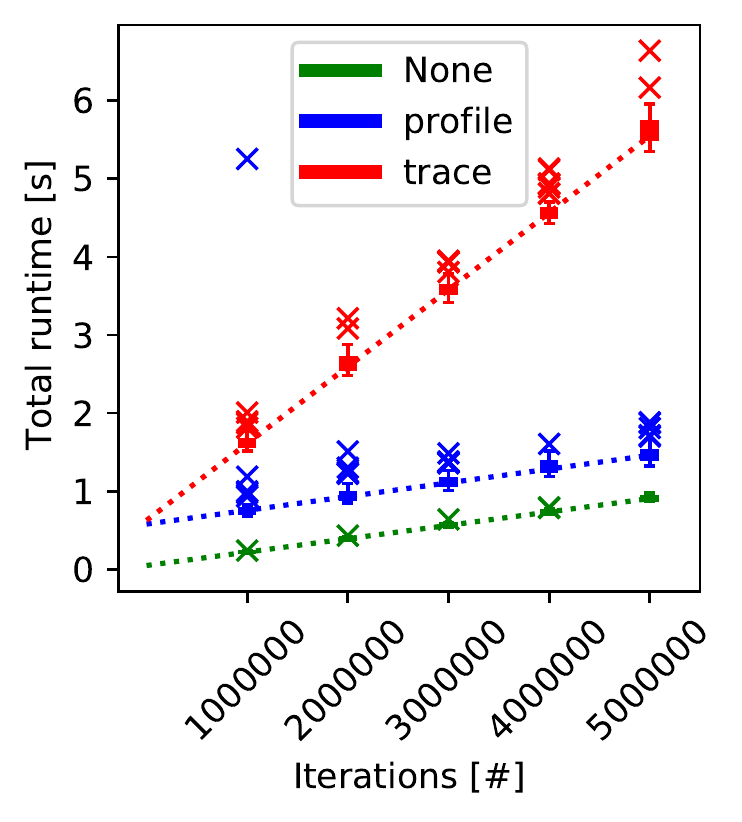}}
	\hfill
	\subfloat[Test case 2 (\label{fig:test_case_2}\lstref{test_2})]{\includegraphics[width=.49\columnwidth,valign=t]{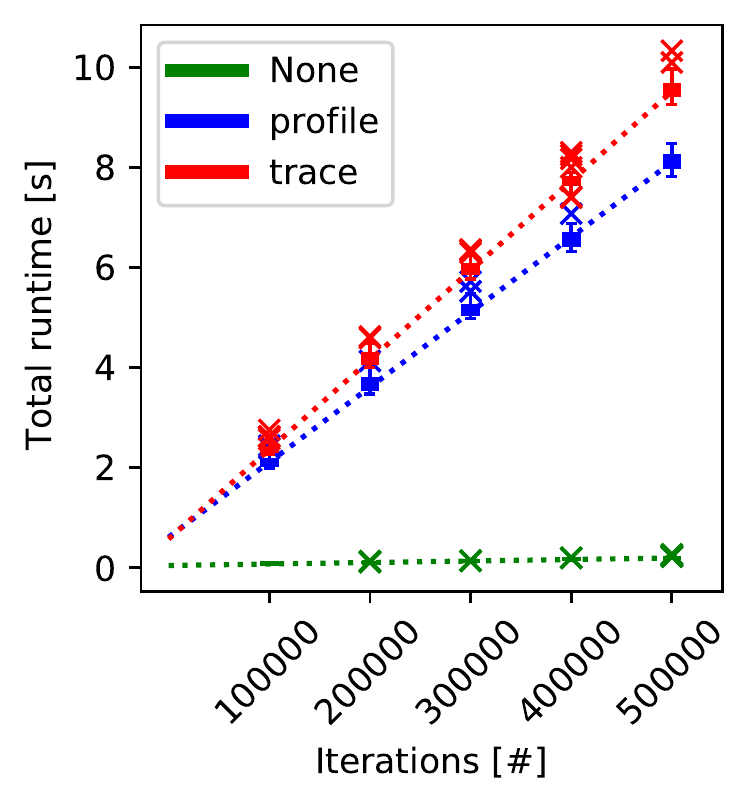}\vphantom{\includegraphics[width=.49\columnwidth,valign=t]{case1.pdf}}}
	\caption{\label{fig:test} Runtime of three instrumenters and non-instrumented code (\none) for different test cases. Dotted lines represent a linear interpolation of the medians of each measurement point.
		The overhead for setting up the measurement and starting the Python environment is 0.6 seconds and independent of the instrumenter. Please note the different x-axis.}
\end{figure}

We performed our experiments on the Haswell partition of the Taurus Cluster at TU Dresden.
Each node is equipped with two Intel Xeon CPU E5-2680 v3 with 12 cores per CPU, and at least 64 GB of main memory per node~\cite{2019_taurus}.
Measurements are taken for each instrumenter, i.e. \setprofile and \settrace, as well as without the Score-P module, marked with \none.
Each experiment is repeated 51 times.
The results are depicted in \figref{test}.
We use linear interpolation to calculate the costs for (a) enabling instrumentation and (b) using the instrumentation.
While the former includes setting up the Python environment and starting and finalizing Score-P, the latter represents the costs to execute one loop iteration.
We disabled the Score-P measurement substrates profiling and tracing to represent only the overhead of instrumenting the code.
The linear interpolation uses the median of each measurement and the \texttt{polyfit} function from numpy to create $t=\alpha + \beta N $ where $t$ represents the runtime, $N$ is the number of iterations, $\alpha$ is the one-time overhead for enabling the instrumentation and $\beta$  is the cost per loop iteration.
The results of this interpolation are presented in \tabref{test_results}.

%


\begin{table}[t]
    \setlength{\tabcolsep}{10pt}
	\centering
	\caption{\label{tab:test_results}Overhead for test cases (median results): $\alpha$: constant overhead; $\beta$: per loop iteration overhead} 
	\begin{tabular}{l|cc|cc}
		& \multicolumn{2}{c|}{Test case 1} & \multicolumn{2}{c}{Test case 2} \\
		Instrumenter       & $\alpha$ & $\beta$ & $\alpha$ & $\beta$ \\ \hline
		\none       & 0.05 s & 0.17 us & 0.05 s & 0.3 us \\
		\setprofile & 0.58 s & 0.18 us & 0.61 s & 15.0 us \\
		\settrace   & 0.63 s & 0.98 us & 0.58 s & 17.9 us \\
	\end{tabular}
\end{table}

For the first test case (\figref{test_case_1}), we see that the instrumentation cost is about \SI{0.6}{\second}.
This cost will apply every time the instrumentation is enabled.
Executing one loop will consume about \SI{0.17}{\micro \second}.
Capturing the loop execution on a per-line scale without forwarding the information to Score-P costs additionally \SI{0.8}{\micro \second}.
This cost only appears for the\settrace instrumenter.

For the second case (\figref{test_case_2}), we see the same initial costs.
However, the per-iteration costs are higher since we call functions.
The general overhead without instrumentation (\none) increases by about \SI{0.13}{\micro \second} to about \SI{0.3}{\micro \second}.
The overhead for function instrumentation increases even more.
Here each function call adds about \SI{14.7}{\micro \second} (for \setprofile).
Due to the per-line overhead, we can say that \settrace should not be used in the current implementation where the same data is given to Score-P by both available instrumenters.
Therefore, we choose to set \setprofile as default instrumenter.
In future versions of our software, we plan to include information on exceptions or executed lines in profiles and traces.
The user will have to choose whether the additional information is important enough for the added overhead.


\section{Related Work}
\label{sec:related}

There are different tools to profile or trace Python code.
The most common ones are the built-in profiling tools \textit{profile} and \textit{cProfile}~\cite{2019_python_profile}.
While both share the same command-line interface, cProfile is preferable, since it is implemented in C and therefore faster.
The output of both tools is usually written to the command line, but can also be re-directed to a file.
The output can be converted and visualised by several third-party tools.
For example, pyprof2calltree~\cite{2019_pyprof2calltree} enables users to convert the output for later analysis with Kcachegrind~\cite{Weidendorfer2008}.
An alternative is SnakeViz~\cite{2019_snakeviz}, which visualises the output of the built-in profilers in a web application.

All these tools are only focussed at single node analysis and do not support parallel programming paradigms used in HPC, like MPI or OpenMP.
This is different for Extrae~\cite{Wagner2017} and TAU~\cite{Shende2006}.
Extrae uses \setprofile{} to register callbacks from Python.
The developers implemented their interface using ctypes, which is a foreign function interface for Python.
TAU version 2.28.1 utilises \texttt{PyEval\_SetProfile} from the C-API and register a callback function that is written in C.

\section{Conclusion and Future Work}
\label{sec:summary}

In this paper, we introduced a module that enables performance engineers to instrument Python applications with \scorep{}.
We described and justified different design decisions that we encountered during development.
To quantify the runtime overhead, we presented measurements of two benchmark kernels.	
Based on these measurements, we decided to use \setprofile as the default instrumenter, as the runtime overhead is smaller than the overhead caused by \settrace.

Further work might include ways to control the runtime overhead, besides manual instrumentation.
One approach could be to sample Python applications.

The Score-P Python bindings are available online at \url{https://github.com/score-p/scorep_binding_python}.


\section*{Acknowledgments}
This work is supported by the European Union’s Horizon 2020 program in the READEX project (grant agreement number 671657).

\bibliographystyle{splncs04} 
\bibliography{paper}

\end{document}